\documentclass[prd,twocolumn,showpacs,amsmath,amssymb]{revtex4}
\usepackage{graphicx}
\usepackage{dcolumn}
\usepackage{bm}

\newcommand{\Dp}{D^{\prime}}
\newcommand{\Fp}{F^{\prime}}

\newcommand{\ag}{a_{3g}}
\newcommand{\aga}{a_{\gamma}}
\newcommand{\ac}{a_c}

\newcommand{\psp}{\psi^{\prime}}

\newcommand{\jpsi}{J/\psi}

\newcommand{\EE}{e^+e^-}

\newcommand{\kskl}{K^0_SK^0_L}

\newcommand{\nnb}{n\overline{n}}

\newcommand{\BBb}{B\overline{B}}

\newcommand{\ppb}{p\overline{p}}

\newcommand{\LLb}{\Lambda \overline{\Lambda}}

\newcommand{\SSbz}{\Sigma^0 \overline{\Sigma}^0}
\newcommand{\SSbp}{\Sigma^+ \overline{\Sigma}^-}
\newcommand{\SSbm}{\Sigma^- \overline{\Sigma}^+}
\newcommand{\XXbz}{\Xi^0 \overline{\Xi}^0}
\newcommand{\XXbm}{\Xi^- \overline{\Xi}^+}
\newcommand{\SzLb}{\Sigma^0 \overline{\Lambda}}

\newcommand{\SbzL}{\overline{\Sigma}^0 \Lambda}
\newcommand{\DDltpp}{\Delta^{++} \overline{\Delta}^{--}}
\newcommand{\DDltp}{\Delta^{+} \overline{\Delta}^{-}}
\newcommand{\DDltz}{\Delta^{0} \overline{\Delta}^{0}}
\newcommand{\DDltn}{\Delta^{-} \overline{\Delta}^{+}}
\newcommand{\OOb}{\Omega^{-} \overline{\Omega}^{+}}

\newcommand{\Heff}{{\cal H}_{eff}}
\newcommand{\Hz}{H_{0}}
\newcommand{\Hmass}{H_{3}^{3}}
\newcommand{\Hchrg}{H_{1}^{1}}
\newcommand{\gz}{g_{0}}

\newcommand{\beq}{\begin{equation}}
\newcommand{\eeq}{\end{equation}}
\newcommand{\beqn}{\begin{eqnarray}}
\newcommand{\eeqn}{\end{eqnarray}}
\newcommand{\beqns}{\begin{eqnarray*}}
\newcommand{\eeqns}{\end{eqnarray*}}
\newcommand{\bfg}{\begin{figure}}
\newcommand{\efg}{\end{figure}}
\newcommand{\bitm}{\begin{itemize}}
\newcommand{\eitm}{\end{itemize}}
\newcommand{\bnum}{\begin{enumerate}}
\newcommand{\enum}{\end{enumerate}}
\newcommand{\btbl}{\begin{table}}
\newcommand{\etbl}{\end{table}}
\newcommand{\btbu}{\begin{tabular}}
\newcommand{\etbu}{\end{tabular}}
\def\eref#1{(\ref{#1})}
\def\Journal#1#2#3#4{{#1} {\bf #2}, #3 (#4)}

\def\NPB{Nucl. Phys. B}

\def\PLB{Phys. Lett. B}

\def\PRL{Phys. Rev. Lett.}
\def\PRD{Phys. Rev. D}

\def\HEPNP{HEP \& NP}

\def\prd#1#2#3 {{~Phys. Rev. D {#1}, #2 (#3) }}  
\def\plb#1#2#3 {{~Phys. Lett. B {#1}, #2 (#3) }}  

\newsavebox{\arrect}
\savebox{\arrect}(0,0){%
\setlength{\unitlength}{0.5cm}
\thicklines
\put(-4,1){\line(1,0){8.0}}
\put(4,1){\line(0,-1){2.0}}
\put(4,-1){\line(-1,0){8.0}}
\put(-4,-1){\line(0,1){2.0}}
\put(0,-1){\vector(0,-1){1.0}}
}

\newsavebox{\arrhomb}
\savebox{\arrhomb}(0,0){%
\setlength{\unitlength}{1cm}
\thicklines
\put(0,0.5){\line(-2,-1){1.0}}
\put(0,0.5){\line(2,-1){1.0}}
\put(0,-0.5){\line(-2,1){1.0}}
\put(0,-0.5){\line(2,1){1.0}}
\put(0,-0.5){\vector(0,-1){0.5}}
\put(-0.15,-0.7){\makebox(0,0)[r]{no}}
\put(1.0,0.0){\vector(1,0){2.0}}
\put(2.0,0.05){\makebox(0,0)[b]{yes}}
}

\newsavebox{\arrparall}
\savebox{\arrparall}(0,0){%
\setlength{\unitlength}{0.5cm}
\thicklines
\put(-3,1){\line(1,0){8.0}}
\put(5,1){\line(-1,-1){2.0}}
\put(3,-1){\line(-1,0){8.0}}
\put(-5,-1){\line(1,1){2.0}}
\put(0,-1){\vector(0,-1){1.0}}
}
\newsavebox{\arrparalla}
\savebox{\arrparalla}(0,0){%
\setlength{\unitlength}{0.5cm}
\thicklines
\put(-3,1){\line(1,0){8.0}}
\put(5,1){\line(-1,-1){2.0}}
\put(3,-1){\line(-1,0){8.0}}
\put(-5,-1){\line(1,1){2.0}}
}

\begin{document}

\title{Symmetry analysis of charmonium decays to two-baryon final state}

\author{X.~H.~Mo$^{1,2}$   J.~Y.~Zhang$^{1}$  
\\  \vspace{0.2cm} {\it
$^{1}$ Institute of High Energy Physics, CAS, Beijing 100049, China\\
$^{2}$ University of Chinese Academy of Sciences, Beijing 100049, China\\
}
}
\email{moxh@ihep.ac.cn}

\date{\today}
	
\begin{abstract}
By virtue of $SU(3)$ flavor symmetry, the effective interaction Hamiltonian is derived in tensor form. The strong and electromagnetic breaking effects are taken into account in a form of ``spurion'' octet, so that the systematical parametrization of all baryon pair final states is realized on basis of the flavor-singlet principle. As an application, the relative phase between the strong and electromagnetic amplitudes is studied in the light of this scenario. In analyzing the data taken at $\jpsi$ resonance region in $\EE$ collider, the details of experimental effects, such as energy spread and initial state radiative correction are taken into consideration in order to obtain the correct results.
\end{abstract}
\maketitle

\section{Introduction}
Since the upgraded Beijing Electron-Positron Collider (BEPCII) and spectrometer detector (BESIII) started data taking in 2008~\cite{bes,yellow}, the largest charm and charmonium data samples in the world were collected, especially the data at $\jpsi$ and $\psp$ resonance peaks, which provide an unprecedented opportunity to acquire useful information for understanding the interaction dynamics of charmonium decays. Although the Standard Model (SM) has been accepted as a universally appreciated theory basis in high energy community, it is still impossible to calculate the wanted experimental observable from the first principle of SM for a great many of processes. The practical evaluations are resorted to the construction of various models, and the powerful tool adopted in actual operation is still the symmetry analysis.

In previous studies, many models are constructed~\cite{Kowalski:1976mc,Haber,a11,zmy2015}, the parametrizations of various decay modes are obtained, such as the pseudoscalar and pseudoscalar mesons (PP), vector and pseudoscalar mesons (VP), octet baryon-pair, and so on. In the present monograph, we focus on the baryon pair final state, and accommodate an approach to acquire the parametrization of different baryon pair final states in a systematical and consistent way.

As an application of the parametrization, the relative phase between the strong and electromagnetic amplitudes of the charmonium decays can be study by virtue of this scenario.

\section{Analysis framework}\label{xct_alsfrk}
In $\EE$ collider experiment, the initial state is obviously flavorless, then the final state should be flavor singlet as well. For charmonium decay, such as $\jpsi$ and $\psp$ decay, final states are hardons composed of light quarks, that is $u, d, s$ quarks. Therefore, the $SU(3)$ group is employed for symmetry analysis. The key principle is that the final state must be flavor-$SU(3)$ singlet. That is among the composition of multiplets, only those containing the singlet are allowed in the effective interaction Hamiltonian. This point will be expound in the following sections. Another crucial issue here is the description of $SU(3)$-symmetry breaking effects, which induced by either the strong interaction or the electromagnetic interaction. Following the suggestion of Ref.~\cite{Haber}, these kinds of effects are treated as ``spurion'' octets. The last but no least, in order to describe final states of both octet and decuplet baryons, the tensor form is adopted to denote the particle multiplet. Although a matrix form is a concise and useful choice, it is only suitable for the octet particles.

\subsection{Parametrization of decuplet-decuplet final state}
We start with decuplet-decuplet baryon final state, and explain in details the notion proposed above. In $SU(3)$ classification, the decuplet contains the isospin multiplets $I=0,\frac{1}{2},1,$ and $\frac{3}{2}$ corresponding respectively to the tensor components $B^{333},B^{i33},B^{ij3},$ and $B^{ijk}$, for $i,j,k=1,2$. These are assigned to the lowest excited baryon states~\cite{quangpham}:
\beq
\left.\begin{array}{llll}
B^{111}=\Delta^{++}&B^{112}=\frac{1}{\sqrt{3}}\Delta^{+}&B^{122}=\frac{1}{\sqrt{3}}\Delta^{0}&B^{222}=\Delta^{-}\\
B^{113}=\frac{1}{\sqrt{3}}\Sigma^{+}&B^{123}=\frac{1}{\sqrt{6}}\Sigma^{0}& B^{223}=\frac{1}{\sqrt{3}}\Sigma^{-}&     \\
B^{133}=\frac{1}{\sqrt{3}}\Xi^{0}&  B^{233}=\frac{1}{\sqrt{3}}\Xi^{-}&  &    \\
B^{333}=\Omega^-     &             &
\end{array}\right.
\label{dkpbyn}
\eeq
The related anti-baryon is denoted as $B_{ijk}$, that is $B_{ijk}=\overline{B}^{ijk}$. It also should be noted that $\Sigma$ and $\Xi$ in decuplet are conventionally denoted as ${\Sigma^*}$ and ${\Xi^*}$ to indicate the excited states, but the star in superscript is suppressed in this subsection without ambiguity. However, when discussing the decuplet-octet final state, the symbol will be recovered to avoid confusion.

According to group theory, the product of two decuplets can be reduced as follows
\beq
{\mathbf 10} \otimes {\mathbf 10^*} = {\mathbf 1} \oplus {\mathbf 8} \oplus {\mathbf 27} \oplus {\mathbf
64}~,
\label{twotenrdn}
\eeq
where the singlet ${\mathbf 1}$ is presented. Therefore, in decuplet-decuplet final state there is a symmetry conserved interaction that can be expressed as
\beq
\Hz = \gz \cdot B_{ijk}B^{ijk}.
\eeq
Here Einstein summation convention is adopted, that is the repeated suffix, once as a subscript and once as a superscript, implies the summation.

Now turn to the question of $SU(3)$-breaking effects. Two types of $SU(3)$ breaking are to be considered. First, the $SU(2)$ isospin symmetry is assumed, that is $m_u=m_d$; but $m_s \neq m_u,m_d$ and this mass difference between $s$ and $u$/$d$ quarks leads to $SU(3)$ breaking. By writing the quark mass term as
$$ m_d(\overline{d}d+\overline{u}u)+m_s \overline{s}s =m_0 \overline{q}q + \frac{1}{\sqrt{3}} (m_d-m_s) \overline{q}
\lambda_8 q~,$$ where $q=(u,d,s)$; $m_0=(2m_d+m_s)/3$ is the average quark mass; $\lambda_8$ is the 8-th Gell-Mann matrix. Explicitly, the matrix ${\mathbf S}_{m}$ is introduced to describe such mass breaking effect
 \beq
 {\mathbf S}_{m}= \frac{g_m}{3} \left(\begin{array}{ccc}
1 &   &  \\
  & 1 &  \\
  &   & -2
\end{array}\right)~,
\label{smbrkmass}
\eeq
where $g_m$ is effective coupling constance due to the mass difference effect.

Second, the electromagnetic effect violates $SU(3)$ invariance since the photon coupling to quarks is proportional to the electric charge:
$$ \frac{2}{3} \overline{u} \gamma_{\mu} u -
\frac{1}{3} \overline{d} \gamma_{\mu} d - \frac{1}{3} \overline{s}
\gamma_{\mu} s = \frac{1}{2} \overline{q} \gamma_{\mu} \left[
\lambda_3+\frac{\lambda_8}{\sqrt{3}} \right] q~.$$ The above
expression indicates that the electromagnetic breaking can be
simulated by the matrix ${\mathbf S}_{e}$ as follows
\beq
{\mathbf S}_{e}= \frac{g_e}{3}
\left(\begin{array}{ccc}
2 &    &  \\
  & -1 &  \\
  &    & -1
\end{array}\right)~,
\label{smbrkcharge}
 \eeq
where $g_e$ is effective coupling constance due to the charge difference effect.

The above discussion is preformed in $SU(3)$ fundamental representation. It is well-known that octet hadron, meson and/or baryon can be expressed by Gell-Mann matrices as well. By virtue of Eqs.~\eref{smbrkmass} and \eref{smbrkcharge}, it is inspiring to consider the $SU(3)$-breaking as one kind of octet. Following the recipe suggested in Ref.~\cite{Haber}, these kinds of $SU(3)$-breaking effects are called ``spurion'' octets.

Now the problem is how to express such effects in tensor form. Again in the light of Eqs.~\eref{smbrkmass} and \eref{smbrkcharge}, it is noticed that ${\mathbf S}_{m}$ is actually the $I$-spin conserved breaking while ${\mathbf S}_{e}$ $U$-spin conserved breaking. Or more physically speaking, ${\mathbf S}_{m}$ is isospin conserved breaking while ${\mathbf S}_{e}$  charge conserved breaking. Therefore, in tensor form, for $I$-spin conserved breaking, it is equivalent to contract up and down indexes along 3 and 3 direction to obtained the effect interaction $\Hmass$, while for $U$-spin conserved breaking, to contract up and down indexes along 1 and 1 direction to obtained the effect interaction $\Hchrg$. Then the effective interaction Hamiltonian reads
\beq
\Heff = \gz \cdot B_{ijk}B^{ijk}+ g_m \cdot \Hmass + g_e \cdot \Hchrg~,
\label{effhmtdkp}
\eeq
where
\beq
\Hchrg =B_{1jk}B^{1jk} -\frac{1}{3} (B_{ijk}B^{ijk})~,
\label{effhmt11}
\eeq
and
\beq
\Hmass =B_{3jk}B^{3jk} -\frac{1}{3} (B_{ijk}B^{ijk})~.
\label{effhmt33}
\eeq

More explanation is in order here. As indicated in Eq.~\eref{twotenrdn}, one octet exists in the reduction of the product of two decuplets. The important fact is that when we treat the breaking effect as an octet as well, in the  reduction of the product of two octets, the singlet exists, as indicated below
\beq
{\mathbf 8} \otimes {\mathbf 8} = {\mathbf 1} \oplus {\mathbf 8} \oplus {\mathbf 8}
\oplus {\mathbf 10} \oplus {\mathbf 10^*} \oplus {\mathbf 27}~.
\label{twoeightrdn}
\eeq
This is the reason that the $\Hmass$  and $\Hchrg$ term can be allowed in $\Heff$. Group theory also indicates that in the reduction of the product of one octet and one ${\mathbf 27}$-tet, or one octet and one ${\mathbf 64}$-tet, there is no singlet, therefore, Eq.~\eref{effhmtdkp} is the final expression of the effective Hamiltonian. The existence of singlet in reduction of multiplet product is the sole criterion to identify which symmetry breaking effect can be allowed in the effective Hamiltonian.

Taking the components of Eq.~\eref{dkpbyn} into the effective Hamiltonian of Eq.~\eref{effhmtdkp}, acquired is the parametrization for decuplet-decuplet baryon final state as listed in Table~\ref{decupletbynform}.

\begin{table}[hbt]
\caption{\label{decupletbynform} Amplitude parametrization forms for decays of a resonance 
into a pair of decuplet baryons, in terms of singlet $A$ (by definition $A=\gz$), as well as charge-breaking term $D$ ($D=g_e/3$) and mass-breaking terms $\Dp$ ($\Dp=g_m/3$).} \center
\begin{tabular}{ll}\hline \hline
  Final state    & Amplitude form  \\ \hline
  $\DDltpp$      & $A+2D-\Dp$    \\
  $\DDltp$       & $A+D-\Dp$     \\
  $\DDltz$       & $A~~~~~~-\Dp$    \\
  $\DDltn$       & $A-D-\Dp$     \\
  $\SSbp$        & $A+D$       \\
  $\SSbz$        & $A$         \\
  $\SSbm$        & $A-D$       \\
  $\XXbz$        & $A~~~~~~+\Dp$     \\
  $\XXbm$        & $A-D+\Dp$    \\
  $\OOb$         & $A-D+2\Dp$         \\
\hline \hline
\end{tabular}
\end{table}

\subsection{Parametrization of octet-octet final state}
The $SU(3)$ octet baryons are convenient to expressed in the matrix notations
\beq
{\mathbf B}=
\left(\begin{array}{ccc}
\Sigma^0/\sqrt{2}+\Lambda/\sqrt{6} & \Sigma^+   & p    \\
\Sigma^-  & -\Sigma^0/\sqrt{2}+\Lambda/\sqrt{6} & n    \\
\Xi^-     & \Xi^0            & -2\Lambda/\sqrt{6}
\end{array}\right)~~,
\label{oktbyn}
\eeq
and
\beq
\overline{\mathbf B}=
\left(\begin{array}{ccc}
\overline{\Sigma}^0/\sqrt{2}+\overline{\Lambda/}\sqrt{6}
                    & \overline{\Sigma}^+ & \overline{\Xi}^+    \\
\overline{\Sigma}^- & -\overline{\Sigma}^0/\sqrt{2}+\overline{\Lambda/}\sqrt{6}
                                          & \overline{\Xi}^0    \\
\overline{p}        & \overline{n} & -2\overline{\Lambda}/\sqrt{6}
\end{array}\right)~~.
\label{atoktbyn}
\eeq
The corresponding tensor notations are respectively $B^i_j$ and $\overline{B}^i_j$, where the superscript denotes the row index of matrix and the subscript the column index. According to reduction of Eq.~\eref{twoeightrdn}, the singlet exists which leads to a symmetry conserved interaction, that is
\beq
\Hz = \gz \cdot \overline{B}^i_j B^j_i~~.
\eeq
As far as breaking terms are concerned, the octet-octet final states are more complex than those of decuplet-decuplet ones. By virtue of Eq.~\eref{twoeightrdn} it is noted that there are two types of octet: an antisymmetric, or $f$-type, and  a symmetric, or $d$-type, defined respectively by
\beq
([\overline{B} B]_f )^i_j = \overline{B}^i_k B^k_j -\overline{B}^k_j B^i_k~~,
\eeq
and
\beq
([\overline{B} B]_d )^i_j = \overline{B}^i_k B^k_j +\overline{B}^k_j B^i_k
-\frac{2}{3} \delta^i_j \cdot (\overline{B}^m_n B^n_m~~).
\eeq
Correspondingly, the each breaking term for octet now contains two parts. In addition, no singlet exists in the reduction of the product of one octet and one decuplet, or one octet and one ${\mathbf 27}$-tet, therefore, the final effective interaction Hamiltonian reads

\beq
\left.\begin{array}{rl}
\Heff = & \gz \cdot \overline{B}^i_j B^j_i    \\
        & + g_m \cdot ([\overline{B} B]_f )^3_3  + g_m^{\prime} \cdot ([\overline{B} B]_d )^3_3   \\
        & + g_e \cdot ([\overline{B} B]_f )^1_1  + g_e^{\prime} \cdot ([\overline{B} B]_d )^1_1 ~~~.
\end{array}\right.~~
\label{effhmtctp}
\eeq
Then writing $\Heff$ in particle form, acquired is the parametrization for the octet-octet baryon final state as listed in Table~\ref{octetbynform}.

\begin{table}[hbt]
\caption{\label{octetbynform} Amplitude parametrization forms for
decays of a resonance 
into a pair of octet baryons, in terms of singlet $A$, as well as symmetric and antisymmetric charge-breaking ($D,F$) and mass-breaking terms ($\Dp,\Fp$). Here $A=\gz$, $D=g_e^{\prime}/3$, $F=-g_e$, $\Dp=-g_m^{\prime}/3$, and $\Fp=g_m$, such a choice is due to the consistency with the previous study in Ref.~\cite{zmy2015}.
}\center
\begin{tabular}{ll}\hline \hline
  Final state    & Amplitude form  \\ \hline
  $\ppb$         & $A+D+F-\Dp+\Fp$    \\
  $\nnb$         & $A-2D-\Dp+\Fp$     \\
  $\SSbp$        & $A+D+F+2\Dp$       \\
  $\SSbz$        & $A+D+2\Dp$         \\
  $\SSbm$        & $A+D-F+2\Dp$       \\
  $\XXbz$        & $A-2D-\Dp-\Fp$     \\
  $\XXbm$        & $A+D-F-\Dp-\Fp$    \\
  $\LLb$         & $A-D-2\Dp$         \\
  $\SzLb+\SbzL$  & $\sqrt{3}D$        \\
\hline \hline
\end{tabular}
\end{table}

\begin{table}[hbt]
\caption{\label{okadkbynform} Amplitude parametrization forms for
decays of a resonance 
into a pair of octet baryons, in terms of
the charge-breaking term $D$ ($D=g_e/2\sqrt{3}$) and the mass-breaking term $\Dp$ ($\Dp=g_m/\sqrt{3}$).} \center
\begin{tabular}{ll}\hline \hline
  Final state    & Amplitude form  \\ \hline
  $\overline{\Sigma^*}^- \Sigma^+$         & $-2D +\Dp$    \\
  $\overline{\Sigma^*}^0 \Sigma^0$         & $+D ~-\Dp$    \\
  $\overline{\Sigma^*}^+ \Sigma^-$         & $~~~~~-\Dp$    \\
  $\overline{\Xi^*}^0 \Xi^0$               & $-2D +\Dp$    \\
  $\overline{\Xi^*}^+ \Xi^-$               & $~~~~~-\Dp$    \\
  $\overline{\Delta}^- p$                  & $~2D$    \\
  $\overline{\Delta}^0 n$                  & $~2D$    \\
  $\overline{\Sigma^*}^0 \Lambda$          & $-\sqrt{3}D$    \\
 \hline \hline
\end{tabular}
\end{table}

\subsection{Parametrization of octet-decuplet final state}
According to the reduction
\beq
{\mathbf 8} \otimes {\mathbf 10^*} = {\mathbf 8} \oplus {\mathbf 10} \oplus {\mathbf 27}
\oplus {\mathbf 35} ~,
\label{teneightrdn}
\eeq
no singlet exits, so there is no symmetry conserved term in the effective interaction Hamiltonian. All terms come from the breaking effects. The octet in Eq.~\eref{teneightrdn} is constructed as follows
\beq
O^i_j = \epsilon^{imn} B_{lmj} B^l_n~~,
\eeq
where $\epsilon^{imn}$ is totally antisymmetric tensor. Since no singlet exists in the reduction of product of a octet with a decuplet, or a ${\mathbf 27}$-tet, or a ${\mathbf 35}$-tet, the sole singlet comes form the product of two octets, and the final effective interaction Hamiltonian reads
\beq
\Heff= g_m O^3_3+ g_e O^1_1~~.
\label{effhmtokdk}
\eeq
The parametrization for octet-decuplet baryon final state is presented in Table~\ref{okadkbynform}.

\section{Experimental measurement}\label{xct_expmsm}
Studying the relative phase between the electromagnetic (EM) and strong decay amplitudes, in addition to the magnitudes of them, provides us a new viewpoint to explore the quarkonium decay dynamics. Studies have been carried out for many $\jpsi$ and $\psp$ two-body mesonic decay modes with various spin-parities:
$1^-0^-$~\cite{dm2exp,mk3exp}, $0^-0^-$~\cite{a00,lopez,a11}, and $1^-1^-$~\cite{a11}, and baryon antibaryon pairs~\cite{ann}. These analyses reveal that there exists a relative orthogonal phase between the EM and strong decay amplitudes~\cite{dm2exp,mk3exp,a00,lopez,a11,ann,suzuki}. The parametrization of baryon pair final state greatly facilitates the further study of this phase, and provides a better knowledge of understanding of the quarkonium decay dynamics.

In the following study, the analysis is performed for the data taken at $\jpsi$ resonance region in $\EE$ collider, the important experimental effects such as the initial radiative correction (ISR) and the effect due to energy spread of accelerator have been taken into account carefully.

\subsection{Born section}

For $\EE$ colliding experiments, there is the inevitable continuum amplitude
$$ 
\EE \rightarrow \gamma^* \rightarrow hadrons
$$ 
which may produce the same final state as the resonance decays do. The total Born cross section is therefore
reads~~\cite{rudaz,wymcgam,Wang:2005sk}
\beq
\sigma_{B}(s) =\frac{4\pi \alpha^2}{3s}
   |\ag(s)+\aga(s)+\ac(s)|^2~{\cal P}(s)~,
\label{bornxc}
\eeq
which consists of three kinds of amplitudes correspond to (a) the strong interaction ($\ag(s)$) presumably through three-gluon annihilation, (b) the electromagnetic interaction ($\aga(s)$) through the annihilation of $c\overline{c}$ pair into a virtual photon,
and (c) the electromagnetic interaction ($\ac(s)$) due to one-photon continuum process. The phase space factor ${\cal P}$ is expressed as
\beq
{\cal P} = v (3-v^2)/2~,
~~v\equiv \sqrt{1-\frac{(m_{B1}+m_{\bar{B}2})^2}{s}}~,
\eeq
where $m_{B1}$ and $m_{\bar{B}2}$ are the masses of the baryon and anti-baryon in the final states, and $v$ velocity of baryon in the center-mass-system. For the octet-baryon-pair decay, the amplitudes have the forms:
\beq
\ac(s)=\frac{Y}{s}~,
\label{ampac}
\eeq
\beq
\aga(s)=\frac{3Y\Gamma_{ee}/(\alpha\sqrt{s})}
{s-M^2+iM\Gamma_t}~,
\label{ampap}
\eeq
\beq
\ag(s)=\frac{3X\Gamma_{ee}/(\alpha\sqrt{s})}
{s-M^2+iM\Gamma_t}~,
\label{ampag}
\eeq
where $\sqrt{s}$ is the center of mass energy, $\alpha$ is the QED fine
structure constant; $M$ and $\Gamma_t$ are the mass and the total width of $\jpsi$; $\Gamma_{ee}$ is the partial width to $\EE$. $X$ and $Y$ are the functions of the amplitude parameters $A,D,F,\Dp$, and $\Fp$ listed in Table~\ref{octetbynform}, viz.
\beq
Y=Y(D,F)~,
\label{defy}
\eeq
\beq
X=X(A,\Dp,\Fp) e^{i\phi}~.
\label{defx}
\eeq
The special form of $X$ or $Y$ depends on the decay mode, as examples, for $\ppb$ decay mode, $X=A-\Dp+\Fp$ and $Y=D+F$ while for $\XXbm$ decay mode, $X=A-\Dp-\Fp$ and $Y=D-F$, according to the parametrization forms in Table~\ref{octetbynform}. In principle, the parameters listed in Table~\ref{octetbynform} could be complex arguments, each with a magnitude together with a phase, so there are totally ten parameters which are too many for nine octet-baryon decay modes. To make the following analysis practical, and referring to the analyses of measonic decays, it is assumed that there is not relative phases among the strong-originated amplitudes $A$, $\Dp$, $\Fp$, and electromagnet amplitudes $D$, $F$; the sole phase (denoted by $\phi$ in Eq.~\eref{defx} ) is between the strong and electromagnet interactions, that is between $X$ and $Y$, as indicated in Eqs.~\eref{defx} and \eref{defy}, where $A$, $D$, $F$, $\Dp$, and $\Fp$ are treated actually as real numbers.

\subsection{Observed section}

In $\EE$ collision, the Born order cross section is
modified by the initial state radiation in the way~\cite{rad.1}
\begin{equation}
\sigma_{r.c.} (s)=\int \limits_{0}^{x_m} dx
F(x,s) \frac{\sigma_{Born}(s(1-x))}{|1-\Pi (s(1-x))|^2},
\label{eq_isr}
\end{equation}
where $x_m=1-s'/s$. $F(x,s)$ is the radiative function been calculated to an accuracy of
0.1\%~\cite{rad.1,rad.2,rad.3}, and $\Pi(s)$ is the vacuum polarization factor. In the
upper limit of the integration, $\sqrt{s'}$ is the experimentally required minimum
invariant mass of the final particles. In the following analysis, $x_m=0.2$ is used which
corresponds to invariant mass cut of 3.3~GeV.

By convention, $\Gamma_{ee}$ has the QED vacuum polarization in its definition~\cite{Tsai,Luth}. Here it is natural to extend this convention to the partial widths of other pure electromagnetic decays, that is
\beq
\Gamma_{f} = 2 \tilde{\Gamma}_{ee}\left(\frac{q_{f}}{M}\right)^3
\left|{\cal F} (M^2) \right|^2~,
\label{eq_defgf}
\eeq
where
$$ \tilde{\Gamma}_{ee} \equiv
\frac{\Gamma_{ee}}{|1-\Pi (M^2)|^2}~$$
with vacuum polarization effect included.

The $\EE$ collider has a finite energy resolution which is much wider than the intrinsic width of narrow resonances such as the $\psp$ and $\jpsi$. Such an energy resolution is usually a Gaussian distribution:
$$
G(W,W^{\prime})=\frac{1}{\sqrt{2 \pi} \Delta}
             e^{ -\frac{(W-W^{\prime})^2}{2 {\Delta}^2} },
$$
where $W=\sqrt{s}$ and $\Delta$, a function of the energy, is the standard deviation of the Gaussian distribution. The experimentally observed cross section is the radiative corrected cross section folded with the energy resolution function
\begin{equation}
\sigma_{obs} (W)=\int \limits_{0}^{\infty}
        dW^{\prime} \sigma_{r.c.} (W^{\prime}) G(W^{\prime},W)~.
\label{eq_engsprd}
\end{equation}

Actually as pointed out in Ref.~\cite{wymcgam}, the radiative correction and the energy spread of the collider are two important factors, both of which reduce the height of the resonance and shift the position of the maximum cross section. Although the ISR is the same for all $\EE$ experiments, the energy spread is quite different for different accelerators. Such a subtle effect must be taken into account in data analysis. In the following analysis all data were assumed to be taken at the energy point which yields the maximum inclusive hadron cross sections in stead of the nominal resonance mass~\cite{wymcgam,wymhepnp}. Some experimental details are summarized in Table~\ref{tab_expcdn}, and they are crucial for the data fitting preformed below.

\begin{table*}[bth]
\caption{\label{tab_expcdn}Breakdown of experiment conditions correspond to
different detectors and accelerators. The data with star ($\ast$) is the
equivalent luminosity calculated by relation ${\cal L}=N_{tot}/\sigma_{max}$. }
\begin{ruledtabular}
\begin{tabular}{llccccc}
         &             &C.M. Energy &Data Taking & Maximum   & Total & Integral    \\
Detector & Accelerator &  Spread    &Position\footnote{
\begin{minipage}{13cm}\mbox{}The data taking position is the energy which
yield the maximum inclusive hadronic cross section. \end{minipage} }
                                                 & section   & event & luminosity  \\
         &             & (MeV)      & (GeV)      &  (nb)     &($\times 10^6$)
                                                                     & (pb$^{-1}$) \\ \hline
\\
BESIII$_a$  & BECPII      & 1.112       & 3.097      &  2830     & 225.3  & 79.6   \\
BESIII$_b$  &             &             & 3.097      &  2830     & 1301.6 & 394.7   \\
BES II  & BEPC        &     0.85    & 3.09700    & 3631.8    & 57.7~~& 15.89$\ast$~~~    \\
MARK II  & SPEAR       &   2.40    & 3.09711    & 1429.3    & 1.32  & 0.924$\ast$    \\
 DM  II  & DCI         &   1.98    & 3.09711    & 1702.0    & 8.6~~ & 5.053$\ast$    \\
FENICE   & ADONE       &   1.24    & 3.09704    & 2595.5    & 0.15  & 0.059$\ast$    \\
\end{tabular}
\end{ruledtabular}
\end{table*}

\subsection{Phase from the fit}\label{xct_fsfit}
Chi-square method is employed to fit the experiment data. The estimator is the defined as
\beq
\chi^2= \sum\limits_i
\frac{[N_i - n_i(\vec{\eta})]^2}{(\delta N_i)^2}~,
\label{chisqbb}
\eeq
where $N$ with the corresponding error ($\delta N$) denotes the experimentally measured number of events while $n$ the theoretically calculated number of events~:
\beq
n={\cal L} \cdot \sigma_{obs} \cdot \epsilon~,
\label{eq_defsig}
\eeq
where ${\cal L}$ is integrated luminosity, $\epsilon$ efficiency, and the observed cross section is calculated according to formula~\eref{eq_engsprd}, which contains the parameters to be fit, such as $A$, $D$, $F$, $\Dp$, $\Fp$, and the phase $\phi$. All these parameters are denoted by the parameter vector $\vec{\eta}$ in Eq.~\eref{chisqbb}.

There are lots of measurements for the octet-baryon-pair decay at $\jpsi$ region. However, many of measurements have been performed almost 30 or 40 years ago~\cite{Brandelik:1979hy}-\cite{Bai:1998fu}. The recent experiment results are mainly from BESII~\cite{besbbdka,besbbdkb,besbbdkc} and  BESIII~\cite{Ablikim:2012eu,Ablikim:2012bw,zjybes3xxb,zjybes3lmd,zjybes3} collaborations.
Besides the data from them, the data from MARKII~\cite{mrk2bbdk} and DMII~\cite{dm2bbdka,dm2bbdkb} are adopted, since the numbers of events from these two experiment group are considerable large and the more information of distinctive decay modes are also provided. All data used in this analysis are summarized in Table~\ref{tab_jpsidt}.
\begin{table*}[bth]
\caption{\label{tab_jpsidt}Experimental data of $\jpsi$ decays to octet baryon pair final states. The first uncertainties are statistical, and the second are systematic. }

\begin{ruledtabular}
\begin{tabular}{ccccc}
  Mode  & $N^{obs}$       & Efficiency   & Branching Ratio        & Detector     \\
        &  (peak)         & (\%)         &     ($\times 10^{-4}$) &     \\ \hline
$\ppb$  &$~63316 \pm 281$ &$48.53\pm0.31$&$22.6 \pm 0.1 \pm 1.4 $& BESII~\cite{besbbdka} \\
        &$~~1420 \pm ~46$ &$49.7 \pm1.6 $&$21.6 \pm 0.7 \pm 1.5 $&MARKII~\cite{mrk2bbdk} \\
        &$314651 \pm 561$ &$66.16\pm0.17$&$21.12\pm 0.04\pm 0.31$& BESIII~\cite{Ablikim:2012eu} \\
$\nnb$  &$~35891 \pm 211$ &$7.69\pm0.06 $&$20.7 \pm0.1  \pm 1.7 $&
       BESIII~\cite{Ablikim:2012eu} \\
$\LLb$  &$~~8887 \pm 132$ &$7.55\pm0.11$ &$20.3 \pm0.3  \pm 1.5 $&BESII~\cite{besbbdkb} \\
        &$~~~365 \pm ~19$ &$17.6\pm0.9 $ &$15.8 \pm0.8  \pm 1.9 $&MARKII~\cite{mrk2bbdk} \\
        &$~~1847 \pm ~67$ &$15.6\pm0.57$ &$13.8 \pm0.5  \pm 2.0 $&DMII~\cite{dm2bbdka}  \\
        &$440675 \pm 670$ &$17.30\pm0.03$&$19.43 \pm0.03\pm0.33 $&BESIII~\cite{zjybes3lmd} \\
$\SSbz$ &$~~1779 \pm ~54$ &$2.31\pm0.07$ &$13.3 \pm0.4  \pm 1.1 $&BESII~\cite{besbbdkb} \\
        &$~~~~90 \pm 10 $ &$4.3 \pm 0.4$ &$15.8 \pm1.6  \pm 2.5 $&MARKII~\cite{mrk2bbdk} \\
        &$~~~884 \pm ~30$ &$9.70\pm0.37$ &$10.6 \pm0.4  \pm 2.3 $&DMII~\cite{dm2bbdka}  \\
        &$111026 \pm335 $ &$7.28\pm0.03$ &$11.64\pm0.04 \pm0.23 $&BESIII~\cite{zjybes3lmd} \\
$\SSbp$ &$~~~399 \pm 26 $ &$0.45\pm0.03$ &$15.0 \pm1.0  \pm 2.2 $&BESII~\cite{besbbdkc} \\
$\XXbz$ &$~~~206 \pm 20 $ &$0.29\pm0.03$ &$12.0 \pm1.2  \pm 2.1 $&BESII~\cite{besbbdkc} \\
        &$134846 \pm 437$ &$8.83\pm0.07$ &$11.65\pm 0.04\pm 0.43$&BESIII~\cite{zjybes3} \\
$\XXbm$ &$~~~194 \pm~14$  &$12.9\pm 0.9$ &$11.4 \pm0.8  \pm 2.0 $&MARKII~\cite{mrk2bbdk} \\
        &$~~~132 \pm~11$  &$2.20\pm0.19$ &$7.0  \pm0.6  \pm 1.2 $&DMII~\cite{dm2bbdkb}   \\
        &$42810.7\pm231.0$&$18.40\pm0.04$&$10.40\pm0.06 \pm 0.74$&BESIII~\cite{zjybes3xxb} \\
$\SzLb+\SbzL$
        &$~~~542 \pm  ~32$&$8.02\pm0.65$ & $0.283  \pm 0.023 $   & BESIII~\cite{Ablikim:2012bw}   \\
\end{tabular}
\end{ruledtabular}
\end{table*}

Since there is lack of the details information about each detectors, it is difficult to deal with all data consistently and accurately. To alleviate
the possible inconsistence among the data from different experiment group,
four relative (relative to the quantity of BESII) normalized factors of luminosity
are introduced with the belief that the relative relations of measurements
of each experiment group is more reliable than the corresponding absolute values.

A remark for BESIII data samples is in order here. There are two sets of $\jpsi$ data samples due to BESIII, which were taken separately in 2009 and 2012. The determination of the total numbers of two data samples and relevant details are contained in Refs.~\cite{jpsisp09} and~\cite{jpsisp09a12}. Two relative normalized factors of luminosity
($f_{bes3a}$ for the data in Refs.~\cite{Ablikim:2012eu,Ablikim:2012bw,zjybes3xxb} and $f_{bes3b}$ for the data in Refs.~BESIII~\cite{zjybes3lmd,zjybes3}) are introduced for BESIII samples.

The fitted parameters are listed as follows:
\beq
\begin{array}{rcl}
  \phi  &=&-89.29^\circ\pm 0.91^\circ~,\mbox{ or } ~+90.71^\circ\pm 0.93^\circ~~; \\
       A &=&~~1.683 \pm 0.005~~, \\
     \Dp &=& -0.097 \pm 0.001~~, \\
     \Fp &=&~~0.183 \pm 0.003~~, \\
      D  &=&~~0.099 \pm 0.003~~, \\
      F  &=&~~0.094 \pm 0.033~~; \\
  f_{mk2}&=&~~0.930 \pm 0.024~~, \\
  f_{dm2}&=&~~0.774 \pm 0.020~~, \\
  f_{bes3a}&=&~~0.914 \pm 0.004~~, \\
  f_{bes3b}&=&~~1.184 \pm 0.007~~.
\end{array}
\label{fitjpsia}
\eeq

Here four factors $f_{mk2}$, $f_{dm2}$, $f_{bes3a}$, and $f_{bes3b}$ are used to normalize
the total integral luminosity for experiments at AMRKII, DMII, and BESIII (two sets of samples), respectively. The fit values indicate that the inconsistencies of
these experiments from that of BESII vary from 10\% to 30\%.

The phase determined from $\jpsi \to \BBb$ decays is fairly consist with the
analysis for $\psp \to \kskl$~\cite{besklks1}, where $\phi$ is determine to
be $(-82^\circ\pm 29^\circ)$ or $(+121^\circ\pm 27^\circ)$. Here the solution $-89.29^\circ$ is more favorable for the universal assumption proposed in Ref.~\cite{Wang:2003zx}. The results of Eq.~\eref{fitjpsia} show that for $\jpsi \to \BBb$ decay the $SU(3)$-symmetric
amplitude ($A$) dominates while others is weak at least at one order of magnitude.

It is also noticed that the results obtained here are consistent with those in Ref.~\cite{Ablikim:2012bw}, in which the ``reduced branching ratio'' method~\cite{a11,LopezCastro:1994xw} was applied, and the $\phi$ is determined to be $(+76 \pm 11)^\circ$. Notice with the ``reduced branching method'' the continuum contribution is simply subtracted from the resonance's, then the interference between them has not been considered properly. Also this method can only provide relative strengths of the different amplitudes.

With the EM amplitudes determined from the fit, one can calculate the
continuum production cross sections of all the final states listed in
Table~\ref{tab_jpsidt}. As a byproduct, we predict
\begin{equation}
\begin{array}{l}
 \sigma(\EE\to \ppb)= 3.45 \pm 1.19 \ \mathrm{pb}, \\
 \sigma(\EE\to \nnb)= 3.60 \pm 0.22 \ \mathrm{pb}, \\
 \sigma(\EE\to \SSbp)= 3.03 \pm  1.05 \ \mathrm{pb},  \\
 \sigma(\EE\to \SSbz)= 0.79 \pm  0.05 \ \mathrm{pb},  \\
 \sigma(\EE\to \XXbz)= 2.75 \pm 0.16 \ \mathrm{pb}, \\
 \sigma(\EE\to \LLb)= 0.84 \pm 0.05 \ \mathrm{pb}, \\
 \sigma(\EE\to \SzLb + \SbzL)= 4.88 \pm 0.50 \ \mathrm{pb},
\end{array}
\end{equation}
at a center of mass energy corresponding to the $\jpsi$ mass; while the cross sections of
$\sigma(\EE\to \SSbm) $ and $\sigma(\EE\to \XXbm)$ are about thousand times smaller. These
can be tested with the data samples at the BESIII experiment.

\section{Summary}\label{xct_sum}
Based on the flavor-singlet principle, assuming the flavor symmetry breaking effect as a ``spurion'' $SU(3)$ octet, the effective interaction Hamiltonian is obtained in tensor form for all kinds of baryon pair final states decaying from a charmonium resonance. It is the first time to acquire such a scheme to systematically parameterize various kinds of baryon pair final states in the light of a single and simple principle. Furthermore, the philosophy of symmetry analysis can be easily extended to the meson pair final state. The corresponding study is in the progress. In a word, such a scheme greatly facilitates the further systematical study of charmonium decays at BESIII experiments, and provides a better knowledge of understanding of the quarkonium decay dynamics.

As a concrete example, by virtue of the parametrization scenario given in this monograph, the data taken at $\jpsi$ resonance region in $\EE$ collider are analyzed to measure the relative phase between the strong and electromagnetic amplitudes. In the analysis the details of experimental effects including the energy spread and the initial state radiative correction are taken into consideration. The existence of a nearly orthogonal relative phase is confirmed at high accuracy based on the data fitting results of $\jpsi$ decays into baryon pair final states.

Since the parametrization forms for all kinds of baryon pair final states are available, more experimental data are being collected and classified for the further analysis.

\section*{Acknowledgment}
The authors acknowledge the helpful discussions with Dr. K.~Zhu. 

\end{document}